# Synthesis of High-Quality Graphene through Electrochemical Exfoliation of Graphite in Alkaline Electrolyte

**Prashant Tripathi, Ch. Ravi Prakash Patel, M. A. Shaz\*and O. N. Srivastava\***

Nanoscience Centre, Department of Physics (Centre of Advanced Studies),
Banaras Hindu University, Varanasi-220005, INDIA

**Abstract**

Owing to wide variety of applications of graphene, high-quality and economical way of synthesizing graphene is highly desirable. In this study, we report a cost effective and simple approach to production of high-quality graphene. Here, the synthesis route is based on electrochemical exfoliation of graphite. Instead of using strong acids (which oxidise and damage the geometrical topology of graphene), we have used alkaline solution (KOH dissolved in water) as electrolyte. TEM analysis shows that as prepared graphene has 1-4 layers and large lateral size (up to ~18μm). Raman analysis shows a unique feature. The D peak which is invariably present in graphene is absent here. This suggests that disorder in the graphene sheets synthesized by the present method, is nearly absent. A tentative mechanism of electrochemical exfoliation in alkaline electrolyte is discussed and described.

*Corresponding authors: email- heponsphy@gmail.com, shaz2001in@yahoo.com

Telephone: 0542-2368468

# 1. Introduction

The one-atom thick graphitic, $sp^2$-bonded carbon material: the graphene which is a two-dimensional hexagonal honeycomb like structure; has attracted significant attention. This material possesses various novel properties of interest such as intrinsically superior electrical conductivity ($10^6$ $\Omega^{-1}$ cm$^{-1}$), nearly transparent in visible light (97.7%) [1], high intrinsic carrier mobility ($2.5 \times 10^5$ cm$^2$ V$^{-1}$s$^{-1}$) [2], high specific surface area (2630 m$^2$g$^{-1}$) [3], excellent mechanical strength (Young's modulus > 1 TPa) [4], and high thermal conductivity (above 3000 W mK$^{-1}$) [5]. These properties make graphene as one of the most promising upcoming material which has potential possibility of applications in electronics, photonics and several other fields [6]. In the field of organic electronics, the biggest challenge is to develop an electrode which should be highly conducting, flexible and transparent besides being very robust [7]. Presently, indium tin oxide (ITO) is most widely used for this purpose but it has its own shortcomings such as brittle nature, sensitivity to acidic and basic environments, high surface roughness, and growing cost due to shortage of indium. Many other materials have been tried earlier but without much success [8-12]. Recent studies suggest that in comparison to ITO, graphene is a better material for making transparent and highly conducting electrode [13-15].

Synthesis of graphene has been investigated through various routes such as micromechanical cleavage [16]. This is time consuming and the yield is poor. Electrochemical synthesis is considered as cheaper and greener method potentially capable of mass production of few-layer graphene (FLG). Using electrochemical exfoliation method Liu *et al.* peeled FLG off graphite at anode in ionic liquids [17]. This solution based method makes graphene amenable to be processed into the form of film, on most substrates including flexible type.

Recently, a new one-step process for the synthesis of high-quality FLG has been developed. This method was first reported by Ching-Yuan Su *et al.* in the year 2011 [13]. In this method electrochemical exfoliation of graphite was carried out in a 0.5M $H_2SO_4$ solution using graphite as anode and platinum as cathode. High-quality graphene but with lower yield (up to 8 wt %) was obtained. The graphene sheets thus obtained were processed through dimethylformamide (DMF) and converted in a film like configuration embodying many graphene sheets in close proximity. Recently, Parvez *et al.* have also used electrochemical exfoliation method for the synthesis of graphene. They have obtained comparatively large graphene sheets (lateral size ~10 μm) by tuning the acidity of $H_2SO_4$ solution to 0.1M solution [15]. It is thus clear that the electrolyte plays crucially important role on the yield and quality of electrochemically synthesized graphene.

Keeping the above said aspect in view, we attempted to use a new electrolyte not used so far. We deployed alkaline electrolyte (KOH dissolved in water). To the best of our knowledge this is the first report employing pure alkaline electrolyte. As will be shown below, the electrochemical exfoliation with alkaline electrolyte leads to better yield, larger lateral extent and no disorder.

## 2. Experimental details

Electrochemical exfoliation was done in an electrolysis cell. This cell has graphite foil (Alfa Aesar) as anode and Pt wire as cathode. The graphite foil was attached to a tungsten wire through silver glue; the contact was protected by rapid repair glue so that the silver paste does not get dissolved during reaction. This whole anode assembly was dipped in electrolyte in such a way that graphite is fully immersed in the electrolyte keeping the contact just above the electrolyte surface. Here we have used alkaline electrolyte by dissolving KOH (Fisher Scientific; 85%) in distilled water. The optimum result for the synthesis of high-quality graphene has been obtained with pH ~13. However, to see the effect of variation of pH on the quality of graphene, we have used electrochemical exfoliation with alkaline electrolyte having pH ~12 and pH ~11. One advantage of using pure alkaline electrolyte may be that it will not oxidize the graphitic anode particularly at the edges. The electrochemical exfoliation was carried out by first applying DC voltage of +3V for 100 sec. After this high voltages (+10V and -10V) were applied on the anode at an interval of 5 sec. The application of high voltages on the anode resulted in the gradual exfoliation of graphite through edges. This process takes place for nearly 30 min to appreciably exfoliate the graphite flakes. During the exfoliation there are two types of graphitic flakes formed; one gets sedimented at the bottom which consists of thick graphitic pieces. The second type of graphitic sample floats on the surface of electrolyte. These flakes are nearly transparent and have been found to consist of few layer graphene (FLG).

## 3. Material characterization

Transmission electron microscopy (TEM) measurements were performed with a FEI Technai-20G$^2$ microscope (acceleration voltage = 200 kV). The Raman spectra of the EG were recorded with Horiba Jobin Yuon (HR800) Raman spectrometer model number H 45517 with an argon-ion laser at an excitation wavelength of 514.532 nm. Fourier transform infrared spectroscopy (FTIR) was obtained by using a Perkin Elmer FTIR spectrometer (spectrum 100).

## 4. Results and discussion

Fig. 1a shows the schematic diagram of experimental setup used for electrochemical exfoliation of graphite. The working electrode, counter electrode and electrolyte used in the setup were graphite foil, platinum wire and KOH dissolved in water respectively. In order to wet the graphite and to enter the hydrated OH$^-$ ions between graphite layers, the static bias of +3V was first applied for 100 sec to the working electrode. The high voltage of +10V was then applied which dissociates the graphite into small sheets which further get spread into the solution. After sometime, it was observed that one type of sheets settle down in the bottom of the electrolysis cell. Presumably, these are multilayer graphene with large thickness. To address the problem of thickness and also to reduce the functionalities of graphitic sheets alternating bias voltages (+10V and -10V) were applied until desired amount of graphitic sheets were obtained. The polarity of voltage changed after every 5 sec. On applying the alternating voltages, it was found that most of the graphitic sheets float on the surface of electrolyte (Fig. 1d). Fig. 1b and c are respectively the optical images of graphite foil before and after exfoliation. Further the vacuum filtration technique was applied to filter the graphene sheets. In order to remove the residual products, filtered graphitic sheets were washed repeatedly with distilled water. The powder then obtained was dispersed in DMF solution by ultrasonication for 20 min, to get exfoliated graphene sheets (Fig. 1e). The dispersion was left for 20 hour for the precipitation of thick graphitic flakes. The upper part of the dispersion was used for further characterization.

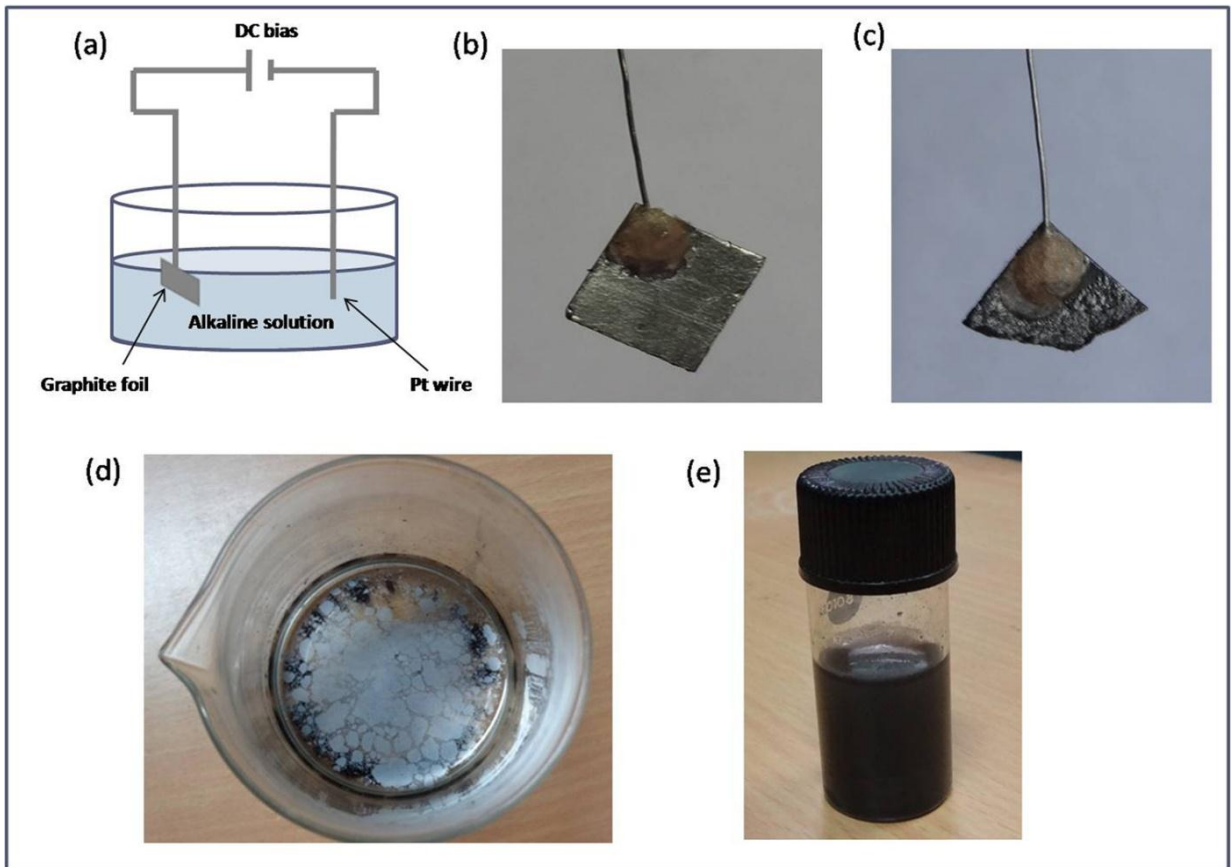

**Fig. 1. (a) Schematic illustration of the electrochemical exfoliation setup. Optical images of (b) graphite foil before exfoliation (c) graphite foil after exfoliation (d) graphene sheets floating on the surface of electrolyte and (e) dispersed graphene sheets in DMF solution.**

One curious aspect observed in the present electrochemical exfoliation was that unlike acidic and ionic liquid electrolytes, the graphite anode did not swell significantly. Another interesting characteristic observed related to the fact that EG appeared like intact graphitic few layer sheets. The damaged edges, wrinkles etc. are often present for graphene produced through other methods. However, in the present method these were almost nonexistent. In fact the exfoliated graphene sheets often exhibited internal morphology of graphite. A typical example is shown in Fig. 2. The graphitic edges bordering graphene are marked.

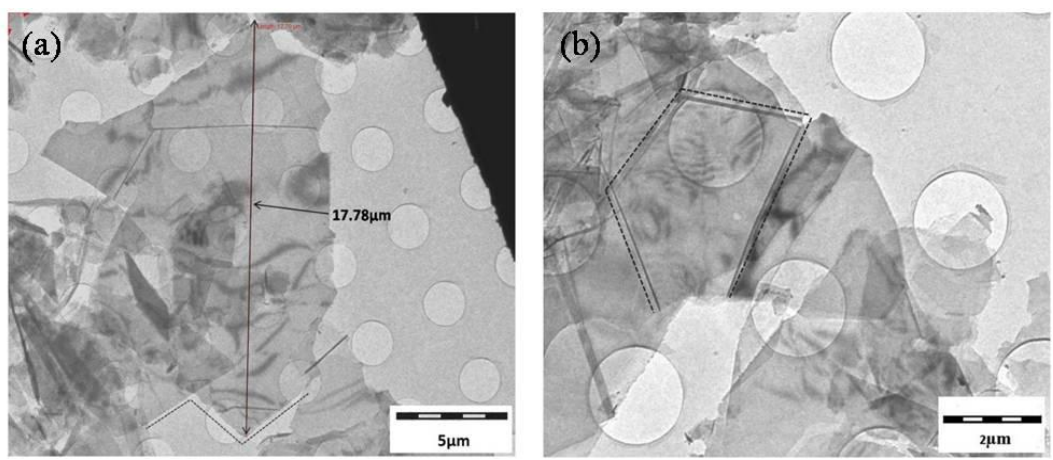

**Fig. 2. The characterization for exfoliated graphene: (a) and (b) the low-magnification TEM images for graphene sheets on carbon holy grid.**

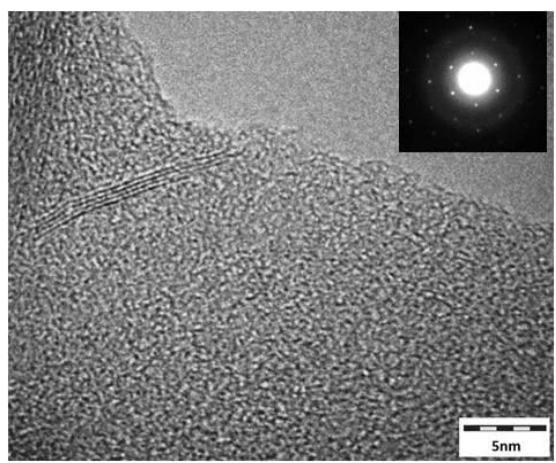

**Fig. 3. HRTEM micrograph of exfoliated four layer graphene.**

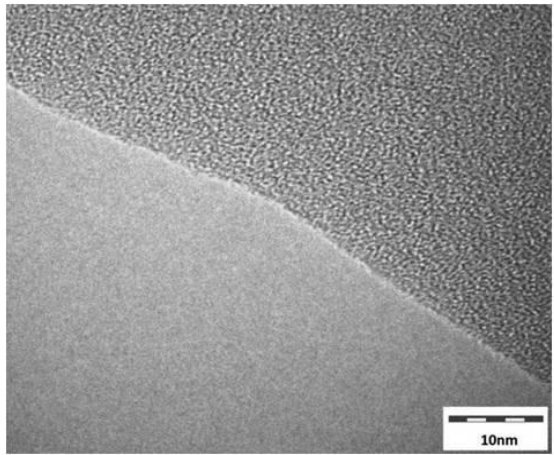

**Fig. 4. HRTEM micrograph of exfoliated graphene (possible single layer graphene)**

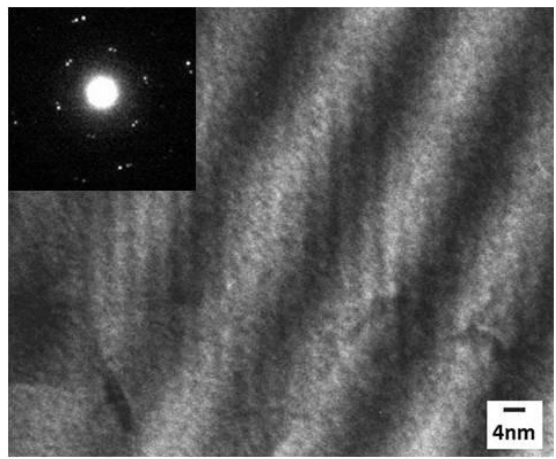

**Fig. 5. Moire patterns in the as synthesized graphene.**

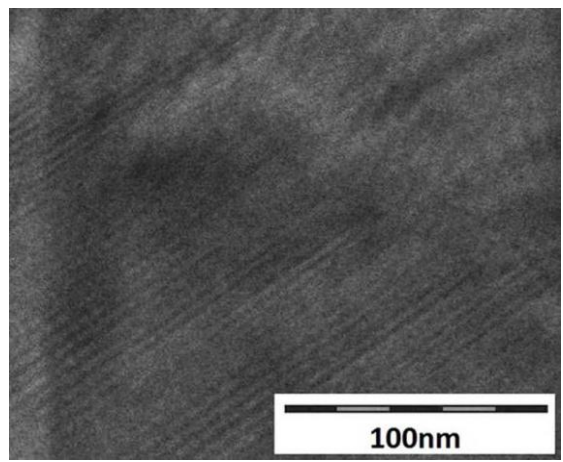

**Fig. 6. Moire pattern of a thick graphene blocks.**

Fig. 2 shows the TEM images of electrochemically exfoliated graphene sheets. From Fig. 2a, it can be concluded that the lateral size of graphene sheets is quite large and upto ~18 μm. The lateral size found in the present study is larger as compared to previously reported lateral sizes of graphene [15,17,18,19]. Fig. 5 shows TEM micrograph exhibiting moire frinzes. The spacing between two broad and small bands were found to be 20.20 nm and 4.27 nm respectively. Fig. 6 is the TEM micrograph of a thick graphene block (showing moire frinzes) bringing out the existence of two spacing 4.67 nm and 9.66 nm.

In order to confirm whether OH$^-$ ions are really responsible for pealing off graphene layers, FTIR spectra were recorded. Fig. 7a shows a representative FTIR spectra of exfoliated graphene sheets. This FTIR spectrum reveals the presence of various functional groups including C-O-C (at 1120 cm$^{-1}$), C-H {at 1344 cm$^{-1}$ and 1377 cm$^{-1}$ (bending), at 2856 cm$^{-1}$ and 2922 cm$^{-1}$ (stretching)}, C=C {at 1625 cm$^{-1}$ (stretching)} and C-O-H {at 3441 cm$^{-1}$ (stretching)} on the graphene samples. The prominent appearance of C-O-H group clearly shows that OH$^-$ radicals did exist in conjunction with exfoliated graphene sheets. We have performed thermal annealing at 450 $^0$C (30 min in 20% H$_2$ and 80% Ar environment). The FTIR spectrum taken after thermal annealing is shown in Fig. 7b. It shows that the intensity of C-O-H and other functional groups decrease indicating the reduction of functional groups as a result of annealing in Ar-H$_2$ mixture.

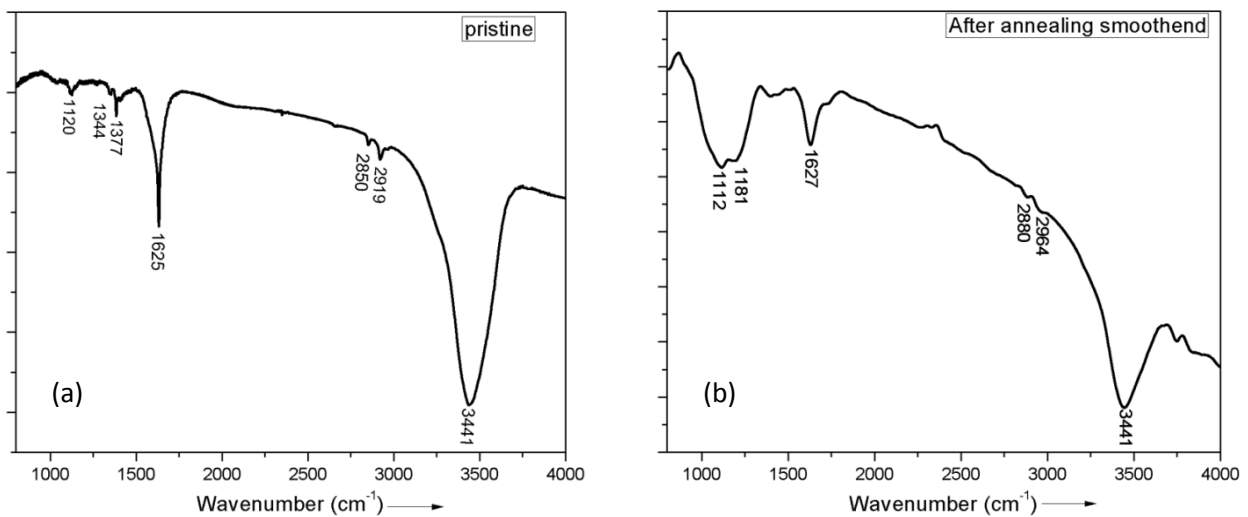

**Fig. 7. FTIR spectrum of graphene sheets (a) and (b) pristine and after annealing at 450 $^o$C in H$_2$/Ar (20sccm/80sccm) atmosphere for 30 min respectively.**

For applications particularly sensor and medical uses, graphene has to be functionalized. Keeping this aspect in view, we have functionalized graphene with amino acid: L-cysteine. The lateral size found in the present case is larger (up to ~18μm) as compared to previously reported lateral size of graphene. This improved lateral size will enhance the loading capacity of ligand. Therefore, in order to maintain the nearly defect free nature of the as synthesized graphene sheet it was functionalized non-covalently with use of amino acid i.e.; L-cysteine. The FTIR spectra recorded confirmed the amine functionalization of the graphene sheet. The interpretation of the result is as follow:

For functionalization 25 mg of graphene sheets were added to 250 ml of double distilled water and then sonicated for one and a half hour for graphene sheets to obtain homogeneous solution. Then, 0.1 M of L-cysteine (double distilled water) was added and sonicated for 30 min followed by 3 h for graphene sheets of constant stirring. The graphene sheets so obtained were thoroughly washed with double distilled water in centrifuge at 10,000 rpm for 10 min and the solution phase was discarded. This washing was repeated 5 times in order to remove any unbounded L-cysteine.

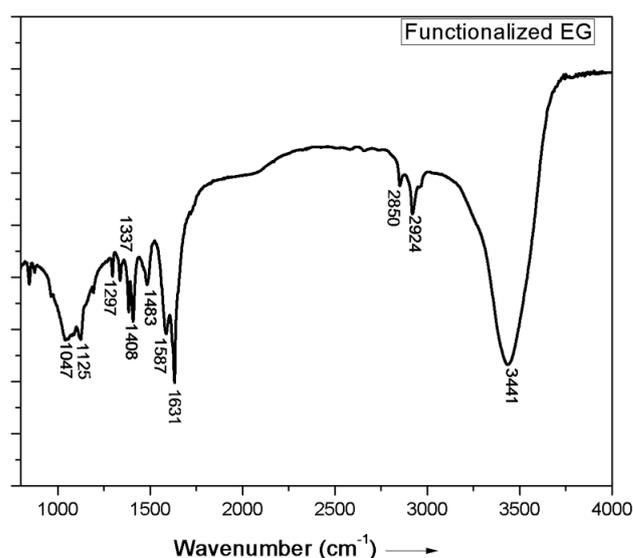

**Fig. 8. FTIR spectrum of functionalized graphene with amino acid: L-cysteine**

In order to test the success of functionalization, FTIR spectra of the functionalized graphenes were recorded. A representative FTIR spectrum of FGS is shown in Fig. 8. Besides the expected absorption peak corresponding to C-O-H at 3441 cm$^{-1}$, other peaks were invariably present. In functionalized graphene sheets an additional peak observed at 2850 cm$^{-1}$ and 2924 cm$^{-1}$ which are the stretching vibrations of alkane group while the presence of peak at 1483 cm$^{-1}$ can be assigned to N-H stretching vibrations of amine group and 1408 cm$^{-1}$ is the C-N stretching vibration. Cysteine on the graphene sheets introduced secondary amide groups as was reflected by the emergence of the characteristic secondary amine peak at about 1587 cm$^{-1}$.

In order to explore the crystallanity of the exfoliated graphene sheets Raman spectra of the graphene were recorded. Fig. 9 shows a typical Raman spectrum. It consists of two peaks in the range of 1500-2800 cm$^{-1}$, particularly at 1580 cm$^{-1}$ and 2730 cm$^{-1}$. The peak at 1580 cm$^{-1}$ is termed as G-peak and arises due to stretching of C-C bond in graphitic material.

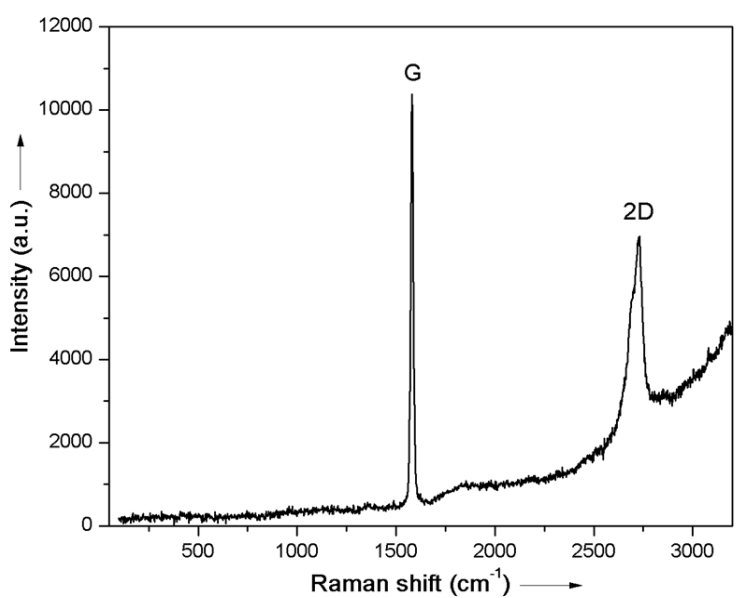

**Fig. 9. Raman spectrum (excited by 514.532 nm laser) of exfoliated graphene sheets.**

Due to its high sensitivity to strain effects in sp$^2$ system, this peak can be used to discuss disorder in graphene sheets. The G peak is quite sharp (FWHM ~15.5 cm$^{-1}$). Also, it is known that larger $I_{2D}/I_G$ (the integrated peak ratio between 2D- and G- bands) shows the higher degree of $sp^2$-hybridized carbon-carbon bonding in graphene structure [20-24]. In the present case $I_{2D}/I_G$ ~3.0 which is high as compare to previous studies [15, 25]. The 2-D peak at 2730 cm$^{-1}$ brings out the signature of graphene. A curious characteristics exhibited by the Raman spectra shown in Fig. 9 is the complete absence of D peak. As is known this peaks originates due to disorder present in the graphene sheets. This peak is invariably present in all the graphene prepared through exfoliation. This also includes electrochemical exfoliation using ionic and acidic electrolytes [13, 15, and 17]. It can thus be said that the present method where basic electrolytes are used, produces nearly defect free graphene sheets. The present method of preparation of graphene sheets in fact corresponds to taking out individual or FLG sheets intact from the graphitic stacks.

## 5. Mechanism of formation of graphitic sheets by electrochemical exfoliation under alkaline electrolyte

Most of the graphene synthesis through electrochemical exfoliation has used acidic electrolyte typified by $H_2SO_4$. The $H_2SO_4$ in solution have $SO_4^{2-}$ ion which has size of ~4.60 Å [26]. This is about ~1.37 times higher than the interlayer spacing of 3.35 Å between graphitic layers in graphite. The $SO_4^{2-}$ ions under applied voltage intercalate between graphitic layers and being bigger (~4.60 Å) they push the graphitic layers farther apart making the spacing greater than 3.35 Å. It leads to weakening of Van der Waals bonding which eventually leads to exfoliation.

In the present electrochemical exfoliation, alkaline electrolyte having preponderance of OH$^-$ ions has been used. The OH$^-$ ions have a size of ~0.958Å [27], which is much smaller than graphitic interlayer spacing of 3.35 Å. Even the hydrated OH$^-$ ions has size of ~2.503 Å [27], which again is smaller than graphitic interlayer spacing of 3.35 Å.

In view of the above, we believe that electrochemical exfoliation mechanism in the present case is different than that observed with acidic and ionic liquid electrolytes. This type of exfoliation is triggered by intercalation of ions like $SO_4^{2-}$ (~4.60 Å) sizes longer than interlayer spacing between graphite layers (3.35 Å). In the following, we proceed to give a tangible mechanism of exfoliation with alkaline electrolyte.

The electrolyte employed by us contains hydrated $OH^-$ ions (size ~2.503 Å). Like water this molecular configuration will be polar. As it enters the graphitic layers under applied electric field, it will polarize the graphitic layers located above and below. Under this condition hydrated $OH^-$ ions will get coupled with the graphitic layers with electrostatic interaction. Under optimum electric force the hydrated ions will pull the graphitic layers above and below out of the graphitic stack. The graphene-hydrated $OH^-$ configuration forming bilayer graphene will thus get formed and float in the alkaline electrolytic solution. The graphene formed by said process will be generally bilayer. However, if one of the layers of the two layers graphene get stack/pinned, while $OH^-$ drags graphite layer single layer graphene can also get formed.

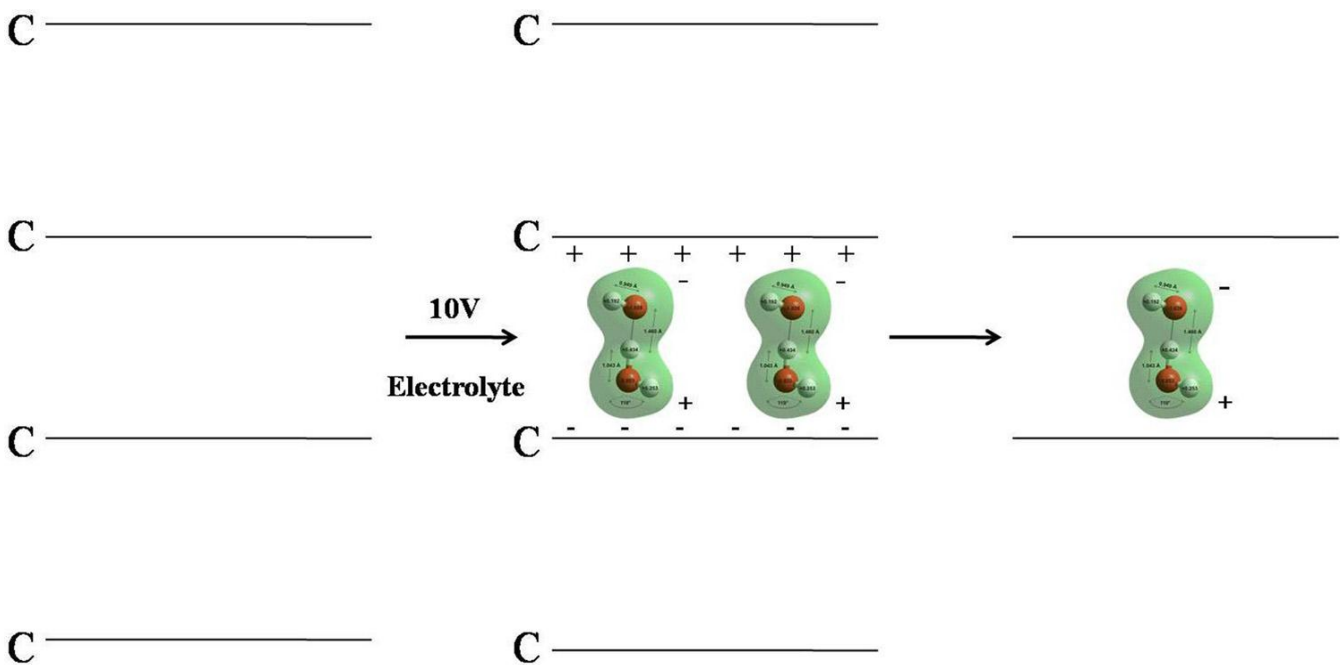

**Fig. 10. Proposed mechanism of electrochemical exfoliation (with alkaline solution) of graphite.**

# 6. Conclusions

In conclusion, we have successfully synthesized graphene by electrochemical exfoliation of graphite employing alkaline electrolyte. The as synthesized graphene has been characterized by TEM, FTIR and Raman found to have 1-4 layer of graphene, contains larger lateral extent and negligible disorder. A feasible mechanism of exfoliation based on polarisation of graphitic layers by $OH^-$ ions has been suggested. The electrostatic interaction of the graphitic layers with $OH^-$ ions which move under the influence of applied electric field leads to the exfoliation. The present study provides a proficient approach to synthesize cost effective and high-quality graphene.

## Acknowledgement:

The authors would like to thanks Dr. Mahe Talat for graphene functionalization studies.